\documentclass[12pt]{amsart}
\usepackage[active]{srcltx}
\usepackage[all
]{xy} \xyoption{poly}
\usepackage{amsfonts,amssymb,amsmath,amsthm,amscd}
\usepackage[a4paper]{geometry}
\usepackage{xspace}
\usepackage{enumitem}
\usepackage[mathscr]{eucal}
\usepackage{amssymb}
\usepackage{a4wide}
\usepackage{ifthen}
\usepackage{amscd}

\def\x{\bf x}
\def\x{\bf x}
\def\y{\bf y}
\def\y{\bf y}

\def\z{\bf z}

\def\U{\bf U}

\newtheorem{thrm}{Theorem}
\newtheorem{thm}{Theorem}[section]
\newtheorem{prop}[thm]{Proposition}
\newtheorem{lem}[thm]{Lemma}
\newtheorem{rem}[thm]{Remark}
\newtheorem{Cor}[thm]{Corollary}

\def\O{\mathbb O}
\def\R{\mathbb R}

\def\C{\mathbb{C}}

\def\R{\mathbb{R}}

\def\Proof{{\it Proof.\ }\ }

\newcommand{\be}{\begin{equation}}
\newcommand{\ee}{\end{equation}}
\newcommand{\bes}{\begin{eqnarray*}}
\newcommand{\ees}{\end{eqnarray*}}
\newcommand{\bee}{\begin{eqnarray}}
\newcommand{\eee}{\end{eqnarray}}

\def\ctd{\hfill$\Box$}

\theoremstyle{remark}

\makeindex

\begin{document}

\begin{center}
{\Large\bf Multi-component generalizations of mKdV equation \\[3mm] and non-associative algebraic structures} \\[7mm]
{\bf Ivan P.\,Shestakov $^{a,b}$,   Vladimir V.\,Sokolov$^{c,d}$}
\end{center} 
 
\bigskip

\begin{minipage}[c]{120mm}
{\small
$a).$Universidade de S\~ao Paulo, S\~ao Paulo, Brazil
 \\[2mm]
$b).$ Sobolev Institute of Mathematics, Novosibirsk, Russia
\\[2mm]
$c).$ Landau Institute for Theoretical Physics,  Chernogolovka,   Russia 
\\[2mm]
$d).$ Universidade Federal do ABC,  S\~ao Paulo, Brazil  \\ 
}
\end{minipage}

\begin{abstract} Relations between triple Jordan systems and integrable multi-component models of the modified Korteveg--de Vries type are established. The most general model is related to a pair consisting of a triple Jordan system and a skew-symmetric bilinear operation. If this operation is a Lie bracket, then we arrive at the Lie-Jordan algebras \cite{GS}. 
\end{abstract}

\pagestyle{plain}

\section{\Large \bf Introduction.}

The existence of higher infinitesimal symmetries  (or a commuting flow) is a foundation stone of the symmetry approach to classification of integrable systems (see, for example,  \cite{ss, mss, miksok}.

A higher (or generalized) infinitesimal symmetry  of evolution equation of the form  
\begin{equation}\label{eveq}
  u_{t}=F(u, u_x, u_{xx}, ... , u_{n}), \qquad n \ge 2, 
\end{equation}
is an evolution equation
\begin{equation}\label{evsym}
u_{\tau}=G(u, u_x,  u_{xx}, \dots , u_m), \qquad m \ge 2, 
\end{equation}
which is compatible  (see Section 2) with~\eqref{eveq}. Here we use the notation 
$$
u_t = \frac{\partial u}{\partial t},  \qquad u_\tau = \frac{\partial u}{\partial \tau},  \qquad  u_x = \frac{\partial u}{\partial x},   \qquad  u_{xx} = \frac{\partial^2 u}{\partial x^2},   \qquad u_i=\frac{\partial^i u}{\partial x^i}.
$$

The symmetry approach provides strong necessary conditions of integrability and allows to find all systems of a prescribed type that could be integrable. To prove the integrability for each of equation found by the necessary conditions, one has to find a Lax representation, auto-B\"acklund transformation or a differential substitution that links the equation with an equation known to be integrable. 

Sergey Svinolupov applied the symmetry approach for constructing multi-component generalizations of known polynomial homogeneous integrable evolution PDEs. By definition, an $N$-component system is a generalization of such an equation if 
\begin{itemize}
\item it has the same polynomial structure;
\item for $N=1$ it coincides with the initial PDE; 
\item it possesses higher symmetries which coincide with symmetries of the initial PDE for $N=1$.
\end{itemize}

 One of the most remarkable observations by S. Svinolupov is the
discovery of the fact that such generalizations
are closely related to the well-known nonassociative algebraic structures  such 
as left-symmetric algebras, Jordan algebras,  triple Jordan systems, etc.
This connection allows one to clarify the nature of known vector and
matrix generalizations (see, for instance \cite{For1,For2,For3}) of classical
scalar integrable equations  and to construct some new examples of this
kind \cite{svsok}.

One of his results is related to integrable multi-component generalizations of the celebrated Korteweg-de Vries equation 
$u_t = u_{xxx}+ 6 u u_x$. This equation was integrated by the inverse scattering method \cite{AblSeg81}. Several algebraic structures are associated with this equation. In particular, the KdV equation has infinitely many infinitesimal symmetries. 

In  paper \cite{svin3} systems of the form 
\begin{equation} \label{kort}
u^i_t=u^i_{xxx}+\sum_{k,j}\, C^i_{jk} \,u^k \, u^j_x, \qquad i,j,k=1,\dots,N
\end{equation}
have been considered. Let us associate with  system \eqref{kort} the $N$-dimensional algebra ${\mathcal A}$ with the structural constants $ C^i_{jk}$. Denote the product in ${\mathcal A}$ by $\circ.$

\begin{thrm}\label{thm3}  {\rm \cite{svin3}}. Suppose that the algebra ${\mathcal A}$ is commutative. Then system 
\eqref{kort} has a polynomial symmetry of order  $m \ge 5$  iff ${\mathcal A}$ is a { \it Jordan algebra}.
\end{thrm}

The original proof of Theorem \ref{thm3} was done  by straightforward computation in terms of structural constants. It turns out 
that the computation can be performed in terms of algebraic operations that define the equation and the symmetry. We demonstrate the corresponding technique in Section 2.

If we do not assume that the algebra ${\mathcal A}$ is commutative, then a generalization of Theorem \ref{thm3} is given by the following  

\begin{thrm}\label{thm4} A system 
\eqref{kort} has a polynomial symmetry of order  $m \ge 5$ iff 
\begin{itemize}
\item The identity $$
(x\cdot y-y\cdot x)\cdot z = 0
$$
holds in $\mathcal A.$  In this case the vector space $I = [\mathcal A,\, \mathcal A]$ is an ideal of $\mathcal A$ and $I^2=0$.
\item The quotient algebra ${\mathcal A}/I$ is a Jordan one.
\end{itemize}
\end{thrm}
In particular, if the algebra $\mathcal A$ is simple then it is Jordan.

We did not find this result in Svinolupov's papers. 

Recall that a system of 
equations (\ref{kort}) is called {\it irreducible} if it cannot be reduced to
a block-triangular form by an appropriate linear transformation
(in the case of a  block-triangular system, the functions
$u^1, \dots, u^M$ \ $(M<N)$ satisfy an autonomous system of form
(\ref{kort})). In paper  \cite{svin3} it was proved that  a system  (\ref{kort}) is irreducible if and only if the corresponding algebra $\mathcal A$ is simple. 
Therefore, by  Theorem \ref{thm4} every irreducible system (\ref{kort}) that has higher symmetries  corresponds to a simple Jordan algebra. 

\smallskip
In this paper we construct multi-component generalizations of the modified KdV equation 
\begin{equation}\label{scmkdv}
u_t=u_{xxx} - 6 u^2 u_x, \qquad   u=u(x, t).
\end{equation}
Equation \eqref{scmkdv} is known to be integrable. In particular, it has 
infinitely many infinitesimal symmetries. The simplest symmetry has order 5 and is given by
\begin{equation}\label{msym}
u_{\tau}=u_{xxxxx} - 10 u^2 u_{xxx} - 40 u u_x u_{xx} - 10 u_x^3 + 30 u^4 u_x. 
\end{equation}

In Section 2 we consider  systems of the form
\begin{equation}\label{jormkdv}
u^i_t=u^i_{xxx}+3\,\sum_{j,k,m} B^i_{jkm} u^j u^m u^k_x, \qquad i,j,k=1,\dots,N,
\end{equation}
The factor 3 is inessential since it can be changed by a scaling of the form $u^i \to {\rm const} \,\, u^i$.
  
There exists the following integrable matrix generalization 
\begin{equation}\label{matmkdv1}
{\bf U}_{t}={\bf U}_3-3 {\bf U}^2 {\bf U}_1-3 {\bf U}_1 {\bf U}^2
\end{equation}
of  equation \eqref{scmkdv}.
 Here ${\bf U}(x,t)$ is a matrix of arbitrary size $m \times m$ and 
$$
{\bf U}_i = \frac{\partial^i {\bf U}}{\partial x^i}.
$$
Written in components of the matrix ${\bf U}$, this system belongs to the class of systems of form \eqref{jormkdv}.
For system \eqref{matmkdv1} we have $N=m^2.$ 

Let $B$ be a  triple system with basis ${\bf e}_1,...,{\bf e}_N,$ such that 
$$
B({\bf e}_j,{\bf e}_k,{\bf e}_m) =\sum_i B^i_{jkm} {\bf e}_i.
$$
If $\, U=\sum_k u^k {\bf e}_k, \,$ then 
the algebraic form of  system (\ref{jormkdv}) is given by
\begin{equation} \label{mkdvsvin}
U_t = U_{xxx} + 3 B(U, U_x, U).
\end{equation}
The triple systems $B(x,y,z)$ such that  $B(x,y,z) = B(z,y,x)$  are in one-to-one correspondence with systems \eqref{jormkdv}. Actually, the triple system $B$ is defined up to a constant factor, which corresponds to the scaling $U\to {\rm const}\, U.$

The main observation, made by Svinolupov \cite{svin2}, is that for any triple Jordan system $J(x,y,z)$ the system \eqref{mkdvsvin}, where $B(x,y,x)=J(x,x,y),$ has infinitesimal symmetries. This statement was not proved in \cite{svin2}. For another relations of integrable models and triple Jordan systems see \cite{svin1}.

In Section 2 we prove the Svinolupov's statement. We also  prove the following converse assertion (see Theorem \ref{thm2}): if a system \eqref{jormkdv} has a fifth order symmetry, then 
\begin{equation}\label{BFF}
B(x,y,z) =  F(x,z,y) + F(z,x,y),
\end{equation}
where $F$ is a triple Jordan system. In terms of $F$ the system \eqref{mkdvsvin} has the form 
\begin{equation} \label{mkdvsvinFF}
U_t = U_{xxx} + 6 F(U, U, U_x).
\end{equation}
Its component form is given by 
\begin{equation}\label{jormkdvF}
u^i_t=u^i_{xxx}+6\,\sum_{j,k,m} F^i_{jkm} u^j u^k u^m_x, \qquad i,j,k,m=1,\dots,N,
\end{equation}
where $F^i_{jkm}$ are the structural constants of the triple Jordan system $F$.  

The class of systems \eqref{jormkdvF} is invariant with respect to the group of linear transformations of the variables $u^1,\dots , u^N$. A system \eqref{jormkdvF} is called {\it reducible} if it can be reduced to a system of the form
\begin{equation}\label{red}
\left\{
\begin{array}{ll}
u^i_t=  u^i_{xxx}+6\,\sum_{j,k,m} F^i_{jkm} u^j u^k u^m_x, \qquad  j,k,m, i=1,\dots, l, \\[2mm]
u^i_t=   u^i_{xxx}+6 \, \sum_{j,k,m} F^i_{jkm} u^j u^k u^m_x, \qquad  j,k,m=1,\dots,N,\ i=l+1,\dots, N.
\end{array}
\right.
\end{equation}

In other words, a system is reducible if it has a subsystem of the same form but of lower dimension. In Section 2 we prove (cf. \cite{svin2}) that a system \eqref{jormkdvF} is irreducible iff the corresponding Jordan triple system $F$ is a simple one.

Since a complete classification of simple Jordan triple systems is known \cite{Zel}, we arrive at a series of irreducible systems of mKdV type.

Our main results are related to the systems of the form
\begin{equation} \label{kdvnew}
u^i_t=u^i_{xxx}+3 A^i_{jk} u^k u^j_{xx}+3 B^i_{jkm} u^j u^m u_x^k, \qquad i,j,k,m=1,\dots,N.
\end{equation}
If all constants $A^i_{jk}$ are equal to zero we arrive at systems of the form \eqref{jormkdv}. In general,   $A^i_{jk}$ and $B^i_{jkm}$ can be regarded as structural constants of an algebra and  a triple system, and therefore \eqref{kdvnew} is related to a pair of algebraic structures. 

If $N=1,$ system \eqref{kdvnew} becomes a scalar equation of the form $u_t=u_{xxx} + 3 a u u_{xx} + 3 b u^2 u_x.$ Using the symmetry approach, one can verify that this equation is integrable only if $a=0.$ However, integrable systems \eqref{kdvnew} more general than \eqref{jormkdv} exist for $N\ge 2.$
An example (see \cite[Section 3.9]{kuper}) of such a system is given by 
\begin{equation}\label{mat2}
{\bf U}_{t}={\bf U}_3+3 {\bf U} {\bf U}_2-3 {\bf U}_2 {\bf U}-6 {\bf U} {\bf U}_1 {\bf U}.
\end{equation}
 Here ${\bf U}(x,t)$ is a matrix of arbitrary size or, more general, an element of free associative algebra generated by ${\bf U}={\bf U}_0,{\bf U}_1,{\bf U}_2,...$. 
It is known that \eqref{mat2} has a symmetry of fifth order. 

Equation \eqref{mat2} can be rewritten in the form
\begin{equation}\label{mat22}
{\bf U}_t =  {\bf U}_3+3 [{\bf U}, {\bf U}_2] + 3\,[{\bf U}, \,[{\bf U},\,{\bf U}_1]]  - 3 {\bf U}^2 {\bf U}_1 - 3 {\bf U}_1 {\bf U}^2.
\end{equation}
Two non-linear terms in the right hand side are defined in terms of the matrix commutator while the others are related to the same triple Jordan  system as in 
\eqref{matmkdv1}. 

Let us write \eqref{kdvnew} in the algebraic form 
\begin{equation}\label{mkdvD}
U_{t}=U_3+3 U U_2+3 \{U,U_1,U\},
\end{equation}
where $x y$ and $ \{x,y,z\}$ are some bilinear and trilinear operations with the structural constants $A^i_{jk}$ and $B^i_{jkm}$.

As the main result of the paper, we prove the following

\begin{thrm}\label{thm5} Suppose that the product $x y$ is skew-symmetric: $x y = - y x.$ System  \eqref{mkdvD} has a symmetry of the form 
\begin{equation}\label{sym5g} \begin{array}{c}
u_{\tau}=U_5+5 A_2(U,U_4)+5 A_3 (U_1,U_3)+5 A_4 (U_2,U_2)+ \\[3mm] \qquad 5 B_2(U,U,U_3)+ 5 B_3(U,U_1,U_2)+ 
5 B_4(U_1,U_1,U_1)+\\[3mm] \qquad  5 C_1(U,U,U, U_2)+5 C_2(U,U_1,U_1, U)+5 D_1(U,U,U,U,U_1)
\end{array}
\end{equation}
iff
\begin{equation}\label{twotr}
\{x, y, z\} =  F(x,z,y) + F(z,x,y) + \frac{1}{2} z (x y) + \frac{1}{2} x (z y),
\end{equation}
where $F$ is a triple Jordan system, and the following identities 
\begin{equation}\label{FId1}
x F(y,z,u) = F(x y, z, u)+F(y, x z, u)+F(y,z, x u)
\end{equation}
and 
\begin{equation}\label{FId2} \begin{array}{c}
z \Big( 2 F(x,y,u)-2 F(x,u,y) - x (u y) \Big) +  {\mathcal J}(u y,x,z) +  \\[2mm] x {\mathcal J}(y,z,u) -  
{\mathcal J}(x y,z,u) -  {\mathcal J}(y,x z,u) - { \mathcal J}(y, z, x u) = 0,
\end{array}
\end{equation}
hold. Here 
\begin{equation}\label{Jac}
{ \mathcal J}\,(x,y,z)=z (x y)+x (y z)+y (z x).
\end{equation}
 \end{thrm}
In the case of zero operation  $x y$ we arrive at the formulas  \eqref{BFF} and \eqref{mkdvsvin}.

 A particular case of an algebraic structure from Theorem \ref{thm5} is  the so-called  {\em Lie-Jordan algebra} which was defined in \cite{GS} as an algebra with a bilinear operation $[x,y]$ and a trilinear operation $T(x,y,z)$ satisfying the identities
\bee
[x,y]&=&-[y,x],\label{LJ1} \\ \
T(x,y,z)&=&T(z,y,x),\label{LJ2} \\ \
[[x,y],z]&=&T(x,y,z)-T(y,x,z),\label{LJ3}\\ \
[t,T(x,y,z)]&=& T([t,x],y,z)+T(x,[t,y],z)+T(x,y,[t,z]), \label{LJ4}\\
T(T(x,y,z),t,v)&=&T(T(x,t,v),y,z)-T(x,T(y,v,t),z)+T(x,y,T(z,t,v))\label{LJ5}.
\eee
It is easy to see that a Lie-Jordan algebra is a Lie algebra with respect to the binary operation $[x,y]$ and is a triple Jordan system with respect to the trilinear operation $T(x,y,z)$ which are interrelated by identities \eqref{LJ3}, \eqref{LJ4}. 
It is also clear that  in any Lie-Jordan algebra the operations $x y = [x,y]$, $F[x,y,z]=\frac{1}{2} T(z,y,x)$ satisfy identities \eqref{FId1} and \eqref{FId2}.

An example of a Lie-Jordan algebra  can obtain from any associative algebra $A$ with an involution *: the set $K(A,*)=\{a\in A\,|\,a*=-a\}$ of skew-symmetric elements of $A$ forms a Lie-Jordan algebra with respect to the operations $[a,b]=ab-ba, \ T(a,b,c)=abc+cba$. It was proved in \cite{GS} that, conversely, every Lie-Jordan algebra is isomorphic to an algebra of this type. 

In Section 2, we prove in details Theorem \ref{thm5} in the
special case of zero binary operation.  

The proof of the general theorem \ref{thm5} given in Section 3 does not actually differ from the proof from Section 2. However,  verifications that some sets of identities are equivalent  become rather cumbersome and we do not give all the formulas, restricted ourself by formulations of the corresponding statements. These statements can be verified by the method of undetermined coefficients (see Section 2) with the help of any computer algebra system. 

In Section 4 we describe simple Lie-Jordan algebras (that  generate irreducible systems \eqref{kdvnew}). Besides matrix algebras that correspond to systems \eqref{mat22},  the Lie-Jordan algebras of skew-symmetric elements of matrix algebras with respect to orthogonal and symplectic involutions appear in the description.

\section{\Large \bf Svinolupov's generalizations of MKdV equation related to triple Jordan systems.}\label{sec2}

\medskip

The modified Korteweg-de Vries equation \eqref{scmkdv}
is one of most celebrated equations integrable by the inverse scattering method \cite{AblSeg81}. This equation possesses infinitely many higher (infinitesimal) symmetries of odd orders.

 For rigorous definition of higher symmetries for equation \eqref{eveq} consider the ring 
${\mathcal F}$ of polynomials that depend of finite number of independent variables $u,u_1, u_{2}, \dots$. 
As usual in differential algebra, we have a principle derivation \begin{equation} \label{DD}
D \stackrel{def}{=}  \sum_{i=0}^\infty u_{i+1} \frac{\partial}{\partial u_i},
\end{equation}
 which generates all independent variables $u_i$ starting from $u_0=u$. We associate with equation \eqref{eveq} the infinite-dimensional vector field   
\begin{equation}
\label{Dt} D_F= \sum_{i=0}^\infty D^{i}(F) \frac{\partial}{\partial u_i},
\end{equation}
This vector field commutes with $D$. We call vector fields of the form \eqref{Dt} {\sl evolutionary}. The set of all evolutionary vector fields forms a Lie algebra over $\C.$
  By definition, the compatibility of~\eqref{eveq} and~\eqref{evsym} means that the vector fields $D_F$ and $D_G$ commute.

Equation (\ref{eveq}), where $F$ is a polynomial, 
 is said to be $\lambda $-{\it homogeneous} of {\it order} $\mu $ if it 
admits the one-parameter group of scaling symmetries
$$(x, \ t, \ u)\longrightarrow (\varepsilon^{-1}x, \ \varepsilon^{-\mu} t, \ \varepsilon^{\lambda} u).$$
For $N$-component systems with unknown variables $u^1,...,u^N$ the corresponding 
scaling group has a similar form
\begin{equation}\label{homo}(x,t,u^1,...,u^N)\longrightarrow (\varepsilon^{-1} x, \ \varepsilon^{-\mu} t, \ 
\varepsilon^{\lambda_1} u^1,...,
\varepsilon^{\lambda_N} u^N).
\end{equation}
Equation (\ref{scmkdv}) is homogeneous with $\mu=3, \lambda=1$ and its simplest symmetry \eqref{msym}
is homogeneous with  $\mu=5, \lambda=1.$ The equations  (\ref{scmkdv}), (\ref{msym})  are also invariant with respect to the discrete involution $u\to -u.$  It was proved in \cite{sw} that if a $\lambda$-homogeneous third order scalar equation with $\lambda \ne 0$ has infinitly many symmetries, then it has a symmetry of fifth order. 

\smallskip

Consider multi-component systems of the form \eqref{jormkdv}. The corresponding 
equations \eqref{mkdvsvin} are homogeneous \eqref{homo} with $\mu=3, \lambda=1$ 
and invariant with respect to the discrete involution $U\to -U.$ Without loss of generality we assume that all polynomial symmetries enjoy the same properties.
Indeed, if \eqref{mkdvsvin} has a polynomial symmetry then any homogeneous component of its right hand side define a symmetry and we may consider only homogeneous symmetries. Similarly, both parts of a polynomial symmetry, symmetric and skew-symmetric under the  $U\to -U,$ are symmetries  separately. That is why we assume that the symmetry does not contain terms of even degrees.

By analogy with the scalar case we are looking for a fifth order symmetry for \eqref{mkdvsvin}.  Under conditions described above such  symmetry is given by \footnote{We put the coefficients 3 and 5 in \eqref{mkdvsvin} and \eqref{symmkdvsvin} to avoid rational numbers in formulas for $B_i$ and $C$.}
\begin{equation}\label{symmkdvsvin}
U_{\tau}=U_{5}+5 B_1(U,U,U_3)+5 B_2(U,U_1,U_2)+5 B_3(U_1,U_1,U_1)+5 C(U,U,U,U,U_1),
\end{equation}
where $B_i$ are some triple systems and $C$ is a 5-system.  

\smallskip

In \cite{svin2} it was formulated the following

\begin{thrm}\label{thm1}  For any  triple Jordan system $[\cdot,\cdot,\cdot]$, the equation \eqref{mkdvsvin}, where $B(x,y,x)=[x,x,y]$  has a fifth order symmetry of the form 
\eqref{symmkdvsvin}. 
\end{thrm}

The original (not published) proof of Theorem \ref{thm1} was done by straightforward computations in terms of structural constants of operations 
$B,B_1,B_2,B_3$   and $C$. It turns out 
that the computations can be performed in terms of these algebraic operations  and identities, which relate them. This drastically simplifies the proof. 
 
\smallskip

In this section we prove a bit stronger statement. 

\begin{thrm}\label{thm2} The equation \eqref{mkdvsvin} has a fifth 
order symmetry of the form \eqref{symmkdvsvin}
iff  
\begin{equation}\label{BBB}
B(x,y,z)=[x,z,y]+[z,x,y],
\end{equation}
where $[\cdot,\cdot,\cdot]$ is a triple Jordan system.
\end{thrm}
{\Proof} The compatibility condition
\begin{equation}\label{com}
0=(U_t)_{\tau}-(U_{\tau})_t\stackrel{def} {=}P(U,U_1,...,U_5)
\end{equation}
of (\ref{mkdvsvin}) and (\ref{symmkdvsvin}) 
leads to a polynomial $P$ that should be identically zero. Thes polynomial has the following structure:
 
$$
\begin{array}{l}
 P = -15\,\Big(\,2 B(U, U_2, U_4) + 4 B(U, U_3, U_3) + 
   4 B(U, U_4, U_2) + 2 B(U, U_5, U_1) +  \\[2mm] \qquad \qquad
   2 B(U_1, U_1, U_4) + 8 B(U_1, U_2, U_3) + 
   12 B(U_1, U_3, U_2) + 2 B(U_1, U_4, U_1) +  \\[2mm] \qquad \qquad
   4 B(U_2, U_1, U_3) + 6 B(U_2, U_2, U_2) - 
   B_1(U, U_1, U_5) - B_1(U, U_2, U_4) -  \\[2mm] \qquad \qquad
   B_1(U_1, U, U_5) - B_1(U_1, U_2, U_3) - 
   B_1(U_2, U, U_4) - B_1(U_2, U_1, U_3) -  \\[2mm] \qquad \qquad
   B_2(U, U_2, U_4) - B_2(U, U_3, U_3) - 
   B_2(U_1, U_1, U_4) - 2 B_2(U_1, U_2, U_3) -  \\[2mm] \qquad \qquad
   B_2(U_1, U_3, U_2) - B_2(U_2, U_1, U_3) - 
   B_2(U_2, U_2, U_2) - B_3(U_1, U_2, U_3) -  \\[2mm] \qquad \qquad
   B_3(U_1, U_3, U_2) - B_3(U_2, U_1, U_3) - 
   2 B_3(U_2, U_2, U_2) - B_3(U_2, U_3, U_1) -  \\[2mm] \qquad \qquad
   B_3(U_3, U_1, U_2) - B_3(U_3, U_2, U_1)\,  \Big) + \dots \, ,
	\end{array}
$$
where dots mean terms of degrees 5 and 7.

After the scaling $U_i \to z_i U_i$ in $P$ all coefficients of different monomials in $z_0,..., z_5$ have to be identically zero. 
Since both system \eqref{mkdvsvin} and symmetry \eqref{symmkdvsvin} are homogeneous, the polynomial $P$ is also homogeneous.  If we assign the weight $i+1$ to $z_i$, then it has the weight 9.

The cubic part of $P$ shown above contains six independent coefficients of $z_1^2 z_4 ,\, z_0  z_3^2,\, z_2^3,\, 
  z_0  z_1  z_5 ,\, z_0  z_2  z_4 , \,
  z_1  z_2  z_3 $, the fifth degree part produces 5 coefficients of $z_0^4 z_4 ,\, z_0^3 z_2^2, \,
  z_0  z_1^4,\, z_0^3 z_1  z_3 , \,
  z_0^2 z_1^2 z_2$ and the seventh degree terms correspond to $z_0^6 z_2,\, z_0^5 z_1^2.$
	
\subsection{Coefficients of symmetry and identities} Equating the coefficient of $z_0 z_1 z_5$ to zero (or, the same, comparing terms with $U, U_1, U_5$), we find that 
\begin{equation} \label{BU1}
2 B(U, U_5, U_1) = B_1(U,U_1,U_5)+B_1(U_1,U,U_5).
\end{equation}
According to \eqref{symmkdvsvin}, we only need to find $B_1(U,U,U_3)$. It follows from \eqref{BU1} that
\begin{equation}\label{B1}
B_1(x, x, y)=B(x, y, x).
\end{equation}
Considering the coefficient of $z_0 z_2 z_4$ and taking into account \eqref{BU1}, we obtain
\begin{equation}\label{B2}
B_2(x, y, z)= 2 B(x, y, z)+2 B(x, z, y).
\end{equation}
Comparing the coefficients of $z_1 z_2 z_3$, we get
\begin{equation}\label{B3}
B_3(x, x, x)= B(x, x, x),
\end{equation}
All other cubic  terms in $P$ disappear by virtue of \eqref{B1}-\eqref{B3}.

Consider now the terms of fifth degree in $P$. The coefficient of $z_3 z_1 z_0^3$ gives rise to 
\begin{equation}\label{CC}
C(x, x, x, x, y)= B(x, B(x,y,x), x) + \frac{1}{2} B(x,y,B(x,x,x)).
\end{equation}
Thus the symmetry \eqref{symmkdvsvin} is expressed in terms of the triple system $B$ and the remaining non-zero coefficients produce identites for $B$. There are only four fifth degree identities $I_i=0, \,i=1,2,3,4$, where 
\bes
I_1(x,y,z)&=&2 B(x,z,B(x,x,y))-3 B(x,z,B(x,y,x))+B(y,z,B(x,x,x)),\\[2mm]
I_2(x,y)&=&2 B(x,y,B(x,x,y))-3 B(x,y,B(x,y,x))+B(y,y,B(x,x,x)),\\[2mm]
I_3(x,y,z)&=&2 B(x, y, B(x, y, z)) - 6 B(x, y, B(x, z, y)) +2 B(x, y, B(y, x, z))   \\
&-&4 B(x, z, B(x, y, y))+2 B(x, z, B(y, x, y)) - 
 2 B(x,  B(y, z, y), x)\\
& +& 2 B(y, y,  B(x, x, z))  -3 B(y, y,  B(x, z, x)) + 4 B(y, z, B(x, x, y))\\
 &-& 2 B(y, z, B(x, y, x)) + 4 B(y, B(x, y, x), z) +  2 B(y, B(x, z, x),y)\\
 &+& 2 B(z, y, B(x, x, y)) -  3 B(z, y, B(x, y, x)),
 \ees
and
\bes
I_4(x,y)&=&B(x, y, B(y, y, y)) + 2 B(y, y, B(x, y, y))\\
 &-& B(y, y, B(y, x, y)) - 
 2 B(y, B(x, y, y), y).
\ees
They come from coefficients of $z_0^3 z_1 z_3, z_0^3 z_2^2, z_0^2 z_1^2 z_2$ and $z_0 z_1^4$, respectively.

\smallskip
In addition to the above fifth degree identities there exist only two identities of degree~7. Namely, the coefficient of $z_0^6 z_2$ in the polynomial $P$ yields the identity
 $$
 B(x, B(x, y, x),\, B(x, x, x)) - B(x, B(x, y, B(x, x, x)), x)=0
 $$
 while the coefficient of $z_0^5 z_1^2$ leads to the identity 
 \bes
 &&2 B(x, y, B(x, x, B(x, y, x))) - 2 B(x, y, B(x, y, B(x, x, x))) \\
 &-&  3 B(x, y, B(x, B(x, y, x), x)) +
  4 B(x, B(x, y, x), B(x, y, x)) \\
  &+& 
 2 B(x, B(x, y, y), B(x, x, x)) - 2 B(x, B(x, y, B(x, x, y)), x) \\
 &+&3 B(x, B(x, y, B(x, y, x)), x) - 4 B(x, B(x, B(x, y, x), y), x)  \\
 &-&B(x, B(y, y, B(x, x, x)), x) + B(B(x, x, x), y, B(x, y, x))=0.
 \ees

\medskip

\subsection{Equivalence of identities} The next part of the proof is an identity handling. 
It is clear that $I_2(x,y)=I_1(x,y,y).$ Using  the method of undetermined coefficients, we will show that the identity $I_4=0$ is a consequence of the identities $I_1=0$ and $I_3=0.$ First, we introduce the polarizations of these identities. 
Let \begin{itemize} \item $J_1(x,y,z,u,v)$ be the coefficient of $k_1 k_2 k_3 $ in $I_1(k_1 x + k_2 u + k_3 v, y, z)$; \\
\item $J_3(x,y,z,u,v)$ be the coefficient of $k_1 k_2 k_3 k_4$ in $I_3(k_1 x + k_2 u, k_3 y + k_4 v, z)$; \\
\item $J_4(x,y,z,u,v)$ be the coefficient of $k_1 k_2 k_3 k_4$ in $I_4(x, k_1 y + k_2 z + k_3 u + k_4 v)$.
\end{itemize}
 Consider the following expression
 $$
 \begin{array}{c}
 Z=J_4(x,y,z,u,v)-\sum_{\sigma\in S_5} b_{\sigma} J_1\Big(\sigma(x),\sigma(y),\sigma(z),\sigma(u),\sigma(v)\Big)-\\[2mm]
  \sum_{\sigma\in S_5} c_{\sigma} J_3\Big(\sigma(x),\sigma(y),\sigma(z),\sigma(u),\sigma(v)\Big),
 \end{array}
 $$
 where $\sigma$ is a permutation of the set $\{x,y,z,u,v\}.$ To take into account the identity $B(x,y,z)=B(z,y,x),$ we fix the ordering
 $$
 u<v<x<y<z< B(\cdot,\cdot,\cdot)
 $$
 and replace all expressions of the form $B(p,q,r)$ by  $B(r,q,p)$ if $p>r.$ After that, equating the coefficients of similar terms in the relation $Z=0$, we obtain an overdetermined system of linear equations for the coefficients $b_{\sigma}$ and $c_{\sigma}$. Solving this system, we find that
\bes
&& J_4(x,y,z,u,v)=\\
&=&\frac{1}{6}\Big(J_1(u, x, v, y, z) + J_1(u, x, y, v, z) + J_1(u, x, z, v, y) + 
 J_1(v, x, u, y, z) \\
&-& J_3(u, v, x, y, z)-J_3(u, v, x, z, y)-J_3(u, v, y, x, z)- J_3(u, v, z, x, y)\\
&-& J_3(u, y, v, x, z)-J_3(u, y, x, v, z)-J_3(v, u, x, y, z)-J_3(v, u, x, z, y)\\
&-& J_3(v, u, y, x, z) - J_3(v, u, z, x, y)- J_3(v, y, u, x, z) - J_3(x, u, v, y, z)\\
 &-& J_3(x, u, v, z, y)- J_3(x, u, y, z, v) - J_3(x, u, z, y, v)  - J_3(x, v, u, y, z)\\
 &-&J_3(x, v, u, z, y) - J_3(y, u, x, z, v)\Big).
 \ees
Since $J_4=0$ follows from $J_1=J_3=0$, we proved that  $I_4=0$ is a consequence from $I_1=I_3=0$.

Consider a triple system 
\bee\label{JTS}
[x,y,z] = \frac{1}{2} \Big( B(y,z,x)+B(y,x,z)-B(x,y,z)  \Big).
\eee
Using the symmetry of $B(x,y,z)$ with respect to $x$ and $z$, one can easily verified  that
\begin{equation}\label{BB}
B(x,y,z)= [x,z,y] + [z,x,y].
\end{equation}

\begin{lem}\label{lem1}  The identities $J_1=J_3=0$ are equivalent to the fact that the triple system $[x,y,z]$ is Jordan, that is, satisfies the Jordan identity
\bes
[x, y, [u, v, z]] - [[x, y, u], v, z] - 
 [u, v, [x, y, z]] + [u,[ y, x, v], z]=0.
\ees
\end{lem}
{\Proof} Let us rewrite the left side of the Jordan identity in terms of the triple system $B$ by means of \eqref{JTS} and obtain the identity ${\mathcal J} = 0$, where
	\[
 \begin{split}&{\mathcal J}(x,y,z,u,v)=\\
-& B(u, v,  B(x, y, z)) +  B(u, v,  B(x, z, y)) + 
  B(u, v,  B(y, x, z)) -  B(u, z,  B(v, x, y)) + \\
& 
  B(u, z,  B(v, y, x)) +  B(u, z,  B(x, v, y)) + 
  B(u,  B(v, x, y), z) -  B(u,  B(v, y, x), z) -  \\
& 
  B(u,  B(x, v, y), z) +  B(u,  B(x, y, z), v) - 
  B(u,  B(x, z, y), v) -  B(u,  B(y, x, z), v) + \\
& 
  B(v, u,  B(x, y, z)) -  B(v, u,  B(x, z, y)) - 
  B(v, u,  B(y, x, z)) -  B(v, z,  B(u, x, y)) +  \\
& 
  B(v, z,  B(u, y, x)) -  B(v, z,  B(x, u, y)) - 
  B(v,  B(u, x, y), z) +  B(v,  B(u, y, x), z) - \\
& 
  B(v,  B(x, u, y), z) +  B(x, y,  B(u, v, z)) - 
  B(x, y,  B(u, z, v)) -  B(x, y,  B(v, u, z)) -  \\
& 
  B(x,  B(u, v, z), y) +  B(x,  B(u, z, v), y) + 
  B(x,  B(v, u, z), y) -  B(y, x,  B(u, v, z)) + \\
& 
  B(y, x,  B(u, z, v)) +  B(y, x,  B(v, u, z)) - 
  B(z, u,  B(v, x, y)) +  B(z, u,  B(v, y, x)) + \\
&  
  B(z, u,  B(x, v, y)) +  B(z, v,  B(u, x, y)) - 
  B(z, v,  B(u, y, x)) +  B(z, v,  B(x, u, y)).
\end{split}
\]
Substituting the expression \eqref{BB}  for $B$  into the identity ${\mathcal J} = 0$, we come back to the Jordan idenity for $[x,y,z].$ 

 By the method of undetermined coefficients one can verify that the identity ${\mathcal J}=0$ follows from $J_1=J_3=0$ and vice versa,
 each of the identities $J_1$ and $J_3=0$ follows from ${\mathcal J}=0.$ For example, 
 $$
 \begin{array}{c}
J_1(x,y,z,u,v) = -{\mathcal J}(u, x, v, z, y) + {\mathcal J}(u, x, y, z, v) - {\mathcal J}(u, y, x, z, v) + \\[1.5mm]
 {\mathcal J}(v, x, y, z, u) - {\mathcal J}(v, y, u, z, x) - {\mathcal J}(v, y, x, z, u) - \\[1.5mm]
 {\mathcal J}(x, u, v, z, y) + {\mathcal J}(x, u, y, z, v) - {\mathcal J}(x, v, u, z, y) + \\[1.5mm]
 {\mathcal J}(y, u, v, z, x) + {\mathcal J}(y, v, u, z, x) + {\mathcal J}(y, v, x, z, u).
 \end{array}
 $$
 The formulas, which express $J_3$ through ${\mathcal J}$ and ${\mathcal J}$ through $J_1, J_3$, are more complicated.

 Using the method of undetermined coefficients, one can check that both seventh degree identities follows from ${\mathcal J}=0$.  Thus we verified that the set of all identities, which are produced by the compatibility condition \eqref{com}, is equivalent to the formulas \eqref{B1}-\eqref{CC} for the symmetry and to the identity ${\mathcal J}=0$.
\ctd
\smallskip
 
\begin{rem} Since equation \eqref{mkdvsvin} is expressed throught $B(x,y,x)$, it follows from \eqref{BB} that all equations that have the fifth order symmetry are described by Theorem \ref{thm1}.
\end{rem}
\smallskip

\subsection{Irreducible systems}

Let us show that irreducible (see Introduction) systems correspond to simple  
triple systems $B(x,y,z)$ in \eqref{mkdvsvin}.

Recall that a subspace $I$ of a triple system $B$ is called {\em an ideal} if $\{I,B,B\}+\{B,I,B\}+\{B,B,I\}\subseteq I$.

\begin{lem}  \label{lem_2.1}
The system of the form (\ref{jormkdv}) is reducible if and only if  the corresponding triple system $B(x,y,z)$ has a non-trivial ideal $I$. In this case an independent subsystem  corresponds to the quotient-system $B/I$.
\end{lem}
{\Proof}
Assume first that a system $B$ may be reduced to form (\ref{red}). Let $I$ be the subspace spanned by $e_{l+1},\ldots,e_N$. If at least one of the indices $j,k,m$ is more than $l$ then $B_{jkm}^i=0$ for all $i<l$ and $\{e_j,e_k,e_m\}\in I$, which proves that $I$ is an ideal of $B$.

\smallskip
Conversely, assume that $B$ has an ideal $I$. Choose in $B$ a base $e_1,\ldots, e_l,e_{l+1},\ldots,e_N$ such that the last $N-l$ elements form a base of $I$. Let us write $ U=\sum_{k=1}^N u^k {\bf e}_k$ in the form $U=V+W$, where $V=\sum_{k=1}^l u^k {\bf e}_k$ and $W=\sum_{k=l+1}^N u^k {\bf e}_k$, then $W\in I$ and we have
\bes
U_t&=&U_{xxx}+B(U,U_x,U)=V_{xxx}+B(V,V_x,V)\\[1.5mm]
&+&W_{xxx}+B(W,V_x+W_x,V+W)+B(V,W_x,V+W)+B(V,V_x,W).
\ees
We have $B(V,V_x,V)=B(V,V_x,V)|_V+B(V,V_x,V)|_W$,  where $B(V,V_x,V)|_V\in V$, $B(V,V_x,V)|_W\in W$. Now our system is reduced to the subsystems of form (\ref{red})
\bes
V_t&=&V_{xxx}+ B(V,V_x,V)|_V,\\[2mm]
W_t&=&W_{xxx}+ B(V,V_x,V)|_W\\[1.5mm]
&+&B(W,V_x+W_x,V+W)+B(V,W_x,V+W)+B(V,V_x,W).
\ees
It is also clear that the independent subsystem $V$ corresponds to the quotient triple system $B/I$.

\ctd

Since the triple systems  $B(x,y,z$  and $[x,y,z]$ are connected by the invertible polynomial transformation \eqref{BBB}, \eqref{JTS}, the system $B(x,y,z)$ is simple iff (see Corollary \ref{Cor3.9} ) the corresponding Jordan triple system $[x,y,z]$ is simple.

According to the classification of simple Jordan triple systems (see, for example, \cite{Zel}), we may now give the following examples of irreducible integrable  vector mKdV systems admitting fifth order symmetries:

\begin{enumerate}
\item For the triple Jordan system defined on the set of all $m\times m$ matrices by  the operation
\bes
\{\x,\y,\z\}=-\frac{1}{2}(\x\y\z+\z\y\x)
\ees
formula \eqref{mkdvsvinFF} gives the matrix mKdV equation \eqref{matmkdv1}.

\item
Let $V=\R^n$ be a euclidian vector space with the scalar product $(x,y)$. The triple Jordan product on $V$  defined by 
\bes
\{\x,\y,\z\}=(\x,\y)\z+(\z,\y)\x-(\x,\z)\y
\ees
gives a  mKdV system
\bes\label{vecmkdv1}
{\bf u}_t={\bf u}_{xxx}+6 ({\bf u}, {\bf u}) \, {\bf u}_x . 
\ees 
\item
The space of rectangular $m\times n$ matrices $M_{m,n}(\R)$ with  the Jordan product $\{\x,\y,\z\}=\x\y^t\z+\z\y^t\x$ defines a matrix  mKdV system
\bes
\U_t=\U_{xxx}+6 \U\U^t\U_x+6 U_x\U^t\U.
\ees
\item 
The spaces of symmetric (skew-symmetric) $m\times m$ matrices are closed with respect to  triple Jordan matrix product in the item (1) and define mKdV systems of dimensions $m(m+1)/2,\, m(m-1)/2$, respectively.

Similarly, the subspaces of hermitian and skew-hermitian $2m\times 2m$ matrices with respect to symplectic involution define mKdV systems of dimensions $m(2m-1)$ and $2m(m-1)$ respectively. 
\item
The space of $1\times 2$ matrices $M_{1,2}(\O)$ over the algebra of octonions $\O$ with the triple Jordan product $\{\x,\y,\z\}=(\x\y^t)\z$ defines a 16-dimensional mKdV system.
\item
The 27-dimensional exceptional simple Jordan algebra $H(\O_3)$ of $3\times 3$ hermitian matrices over octonions ({\em the Albert algebra}) defines a 27-dimensional mKdV system with respect to the triple Jordan product 
$$
\{\x,\y,\z\}=(\x\cdot \y)\cdot \z+(\z\cdot \y)\cdot \x-(\x\cdot \z)\cdot \y,
$$
 where $\x\cdot \y=\frac12 (\x\y+\y\x)$.
\end{enumerate}

\section{\Large \bf  MKdV type systems related to pairs of compatible algebraic structures}

Conceptually, considerations of this section are very closed to those described in Section 2 although the corresponding computations are more cumbersome. For this reason
we often give only basic outline of proofs. 

We consider homogeneous  mKdV-type systems of the form
\begin{equation}\label{mkdv}
U_{t}=U_3+3 A_1(U,U_2)+3 B_1(U,U_1,U),
\end{equation}
where $A_1(x,y)$ and $B_1(x,y,z)$ are a binary and a ternary operations in the same finite-dimensional vector space $\bf V.$ Without loss of generality we assume that $B_1(x,y,z)=B_1(z,x,y).$
If $\, U=\sum_k u^k {\bf e}_k \,$, where  ${\bf e}_1,...,{\bf e}_N$ is a basis in $\bf V,$ then (\ref{mkdv}) is equivalent to a PDE system of the form \eqref{kdvnew}.
We will use the following notation 
\begin{equation}\label{A1B1}
A_1(x,y)=x y, \qquad B_1(x,y,z)=\{x, y, z\}.
\end{equation}

In this paper we consider the case (cf., \eqref{mat2})
\begin{equation}\label{skew}
x y =-y x.
\end{equation}
\begin{rem} It is possible to prove without any assumptions that the binary operation has to satisfies the following cubic identity
$$(x y +y x) z=0.$$
\end{rem}

The problem is: for which operations $A_1, B_1$ equation (\ref{mkdv}) possesses 
a homogeneous fifth order symmetry. Compairing with \eqref{symmkdvsvin} the symmetry has terms of even degree. The general symmetry ansatz is given by
\begin{equation} \label{sym} \begin{array}{c}
u_{\tau}=U_5+5 A_2(U,U_4)+5 A_3 (U_1,U_3)+5 A_4 (U_2,U_2)+ \\[3mm] \qquad 5 B_2(U,U,U_3)+ 5 B_3(U,U_1,U_2)+ 
5 B_4(U_1,U_1,U_1)+\\[3mm] \qquad  5 C_1(U,U,U, U_2)+5 C_2(U,U_1,U_1, U)+5 D_1(U,U,U,U,U_1).
\end{array}
\end{equation}
The differential polynomial $P$ from \eqref{com} has the form
$$
\begin{array}{c}
P=-15 \, \Big(\,  U_1  U_6 + 2  U_2  U_5 + U_3  U_4 
  - A_2(U_1, U_6) - A_2(U_2, U_5) - \\[2mm] \qquad \qquad
   A_3(U_2, U_5) - A_3(U_3, U_4) - A_4(U_3, U_4) - 
   A_4(U_4, U_3)\, \Big) + \dots \, ,
\end{array}
$$
where the dots symbolize terms up to degree 7. After the scaling $U_i \to z_i U_i$ in $P$ all coefficients at different monomials in $z_0,..., z_6$ have to be identically zero.  As in the previous section, the polynomial $P$ is homogeneous of the weight 9  if we assign the weight $i+1$ to $z_i$.

\subsection{Coefficients of symmetry and fourth degree identities}

Equating the quadratic terms in $P$ to zero, we find that
\begin{equation} \label{AA}
A_2(x,y)=x y, \qquad A_3(x,y)=x y, \qquad A_4(x,x)= 0.
\end{equation}

The vanishing of the cubic terms is equivalent to 
\begin{equation} \label{BBb} \begin{array}{c} 
B_2(x,x,y)=x (x y)+\{x, y, x\}, \\[2mm]
B_3(x,y,z)=x (y z)+(x y) z+
2 \{x, y, z\}+2 \{x, z, y\}, \qquad B_4(x,x,x)=\{x, x, x\}.
\end{array}
\end{equation}
 
The terms of degree 4 give rise to the formulas
\begin{equation} \label{CCc} \begin{array}{c} 
\displaystyle C_1(x,x,x,y)=\frac{1}{3}\Big(-x (x (x y))+3 \{x, x y, x  \} +3 x \{x,y,x \}-y \{x,x,x\}  \Big),
\\[3mm]
\displaystyle C_2(x,y,y,x)=\frac{1}{6}\Big(x (y (x y))+y( x (x y))+2 y  \{x, x, y  \} - 2 y \{x,y,x \}+
\\
12 x  \{x,y,y \}+12  \{x,y,x y \}  \Big).
\end{array}
\end{equation}
Moreover, they produce a  set of 5 fourth degree identites $I_i=0, \,i=1,...,5$, where 

$$
I_1(x,y)=\begin{array}{c}
 x ( x (x y)) -  x \{x, x, y\} + 
  x \{x, y, x\} -  y \{x, x, x\} - 
 3 \{x,  x y, x\},
\end{array}
$$

$$
\begin{array}{c}
I_2(x,y,z)=-x (x (y z)) + 2 x (y (x z)) - 3 x (z (x y)) -  y (x (x z)) 
  - 2 z \{x, x, y\} +\\[2mm]  2 z \{x, y, x\} - 
 3 (x y) (x z),
\end{array}
$$

$$
\begin{array}{c}
I_3(x,y,z) = 2  x (x (y z)) - 5  x ( y (x z)) + 5 x (z (x y)) +  y (x (x z])) -  z (x (x y)) + \\[2mm]
 2  y \{x, z, x\} - 2  y \{x, x, z\} + 
 2  z \{x, x, y\} -  2  z\{x, y, x\} + 
 6 (x y) (x z),  \\[2mm] 
\end{array}
$$

$$
\begin{array}{c}
I_4(x,y,z) = -9  x (y (y z))  + 
 3  y (x (y z))  - z (y (x y)) + 2  y (y (x z)) - y (z (x y)) +  \\[2mm]  12  (x y) (y z) -2 y\{x, y, z\} + 
 4  y\{x, z, y\}- 2  y\{y, x, z\} -  2  z  \{x, y, y\} + \\[2mm] 
 2  z\{y, x, y\}  -  
 12 \{y, z,  x y\}+ 6 x\{y, z, y\} - 6 \{y,  x z, y\},  \\[2mm] 
\end{array}
$$

$$
\begin{array}{c}
I_5(x,y,z) = -5 x(z (y z)) -  y (z (x z)) + z (x (y z))+ 
 5  z (y (x z)) - 8 z (z (x y)) + \\[2mm]
 6 (x z) (y z) +    2  y\{x, z, z\} -
 2  y\{z, x, z\}  + 4 z\{x, y, z\} - 
 2  z\{x, z, y\} -  \\[2mm]  
2  z\{y, x, z\} + 12  x \{y, z, z\} - 12 \{y, z,  x z\} - 
 12 \{y,  x z, z\} - 12 \{z, z,  x y\}.
\end{array}
$$

\begin{prop} The identities $I_1=0, \ldots, I_5=0$ are equivalent to  identities $Q_1=0$, $Q_2=0$, where
\begin{eqnarray}\label{id4-1}
&Q_1(x,y,z,u) = y \{u,z,x\}- \{y u,z,x\}- \{u,y z,x\}- \{u,z,y x\}+&\\
&\frac{1}{2}\Big({ \mathcal J}(u z,x,y)+{ \mathcal J}(x z,u,y)+ u { \mathcal J}(x,y,z)+ x { \mathcal J}(u,y,z)\Big),&\nonumber\\
&Q_2(x,y,z,u) = 2 z \, \Big(\{x,y,u\} - \{x, u,y\} - x (u y) \Big)-&\label{id4-2}\\
&u { \mathcal J}(x,y,z)-y { \mathcal J}(u,x,z)-2 { \mathcal J}(u y,x,z)+ { \mathcal J}(u z,x,y) + { \mathcal J}(y z,u,x).&\nonumber
\end{eqnarray}
Here ${ \mathcal J}$ is defined by \eqref{Jac}.
\end{prop}
{\Proof} Let 
\begin{itemize} 
\item $J_1(x,y,z,u)$ be the coefficient of $k_1 k_2 k_3 $ in $I_1(k_1 x + k_2 z + k_3 u, y)$; \\
\item $J_2(x,y,z,u)$ be the coefficient of $k_1 k_2$ in $I_2(k_1 x + k_2 u, y, z)$; \\
\item $J_3(x,y,z,u)$ be the coefficient of $k_1 k_2$ in $I_3(k_1 x + k_2 u, y, z)$; \\
\item $J_4(x,y,z,u)$ be the coefficient of $k_1 k_2 $ in $I_4(x, k_1 y + k_2 u, z)$; \\
\item $J_5(x,y,z,u)$ be the coefficient of $k_1 k_2  $ in $I_5(x, y, k_1 z + k_2 u )$.
\end{itemize}
\smallskip

  The following formulas show that identities $Q_1=0,\,Q_2=0,\, J_1=0,\,J_2=0,\, J_3=0$ follow from $J_4=0, J_5=0$:
$$
\begin{array}{c}
72\, Q_1(x, y, z, u) = 
  3 J_4(u, y, z, x) + 4 J_4(u, z, x, y) + 3 J_4(u, z, y, x) + 
    J_4(y, u, x, z) + \\[2mm]  3 J_4(y, u, z, x) +J_4(z, u, x, y) + 
    3 J_4(z, u, y, x) - J_5(u, x, z, y)- \\[2mm] 2 J_5(u, y, z, x) - 
    2 J_5(u, z, y, x) + J_5(y, u, z, x) +   2 J_5(y, x, u, z) - \\[2mm]
    2 J_5(y, z, u, x) - 2 J_5(z, u, y, x) - J_5(z, x, u, y) + 
    J_5(z, y, u, x),
\end{array}
$$

$$
\begin{array}{c}
36\, Q_2(x, y, z, u) = -12 J_4(u, y, z, x) - 14 J_4(u, z, x, y) - 12 J_4(u, z, y, x) - 
 2 J_4(y, u, x, z) - \\[2mm] 8 J_4(z, u, x, y) - 9 J_4(z, u, y, x) - 
 3 J_4(z, y, u, x) + 5 J_5(u, x, z, y) + \\[2mm] 7 J_5(u, y, z, x) + 
 7 J_5(u, z, y, x) + J_5(y, u, z, x) - J_5(y, x, u, z) + \\[2mm]
 J_5(y, z, u, x) + 7 J_5(z, u, y, x) + 2 J_5(z, x, u, y) + 
 J_5(z, y, u, x),
\end{array}
$$

$$
\begin{array}{c}
12 J_1(x,y,z,u) = 2 J_5(u, x, z, y) - 4 J_5(u, y, z, x) + 2 J_5(u, z, y, x) + 
 2 J_5(x, u, z, y) -\\[2mm] 4 J_5(x, y, u, z) + 2 J_5(x, z, u, y) - 
 J_5(y, u, z, x) - J_5(y, x, u, z) -\\[2mm] J_5(y, z, u, x) + 
 2 J_5(z, u, y, x) + 2 J_5(z, x, u, y) - 4 J_5(z, y, u, x),
\end{array}
$$

$$
\begin{array}{c}
6 J_2(x,y,z,u) = 6 J_4(u, y, z, x) + 8 J_4(u, z, x, y) + 6 J_4(u, z, y, x) + 
 2 J_4(y, u, x, z) +\\[2mm] 5 J_4(z, u, x, y) + 6 J_4(z, u, y, x) + 
 3 J_4(z, y, u, x) - 2 J_5(u, x, z, y) -\\[2mm] 4 J_5(u, y, z, x) - 
 4 J_5(u, z, y, x) - J_5(y, u, z, x) + J_5(y, x, u, z) - \\[2mm]
 J_5(y, z, u, x) - 4 J_5(z, u, y, x) - 2 J_5(z, x, u, y) - 
 J_5(z, y, u, x),
\end{array}
$$

$$
\begin{array}{c}
6 J_3(x,y,z,u) = 3 J_4(u, y, z, x) + 4 J_4(u, z, x, y) + 3 J_4(u, z, y, x) + 
 J_4(y, u, x, z) -\\[2mm] 3 J_4(y, u, z, x) + 7 J_4(z, u, x, y) + 
 9 J_4(z, u, y, x) + 6 J_4(z, y, u, x) -\\[2mm] J_5(u, x, z, y) - 
 2 J_5(u, y, z, x) - 2 J_5(u, z, y, x) + J_5(y, u, z, x) +\\[2mm] 
 2 J_5(y, x, u, z) - 2 J_5(y, z, u, x) - 5 J_5(z, u, y, x) - 
 4 J_5(z, x, u, y) -\\[2mm]  2 J_5(z, y, u, x).
\end{array}
$$

Vice versa, $J_4=0, J_5=0$ follows from $Q_1=0, Q_2=0$:

$$
\begin{array}{c}
J_4(x,y,z,u) = 2 \Big(6 Q_1(u, x, z, y) - Q_2(u, y, z, x) + 2 Q_2(u, z, y, x) - 
   Q_2(y, u, z, x) +\\[2mm]  2 Q_2(y, z, u, x) - Q_2(z, u, y, x) - 
   Q_2(z, y, u, x)\Big),
	\end{array}
$$

$$
\begin{array}{c}
J_5(x,y,z,u) = 2 \Big(6 Q_1(u, x, z, y) + 6 Q_1(z, x, u, y) + 2 Q_2(u, y, z, x) + 
   Q_2(u, z, y, x) -\\[2mm]  Q_2(y, u, z, x) - Q_2(y, z, u, x) + 
   Q_2(z, u, y, x) + 2 Q_2(z, y, u, x)\Big).
	\end{array}
$$

\begin{rem} The identities $Q_1=Q_2=0$ imply
$$
v \{x,y,z\}+ x \{v,z,y\}+y \{x,v,z\}+z \{v,x,y\} = 0.
$$
\end{rem}

Equating to zero the coefficient of $z_3 z_1 z_0^3$  in $P$, we find that 
\begin{equation} \label{DDd} \begin{array}{c} 
\displaystyle D_1(x,x,x,x,y)=\frac{1}{12}\Big( 6 \{x,y,\{ x,x,x\} \}+12 \{x,\{x,y,x \}, x \}-3 \{x, x (x y), x \} +
\\[3mm]
\displaystyle x \{x,x, x y \}-7  x \{x, x y, x \}-x (x \{x,x,y\})-2 x (x \{x,y,x\})+
\\[3mm]
3 x (y \{x,x,x\})-5 (x y) \{x,x,x\}+4 x (x (x (x y))) \Big).
\end{array}
\end{equation}
\begin{lem} \label{lemma34} Formulas   \eqref{CCc}, \eqref{DDd}  can be simplified to  
 \begin{equation} \begin{array}{c} \label{CCD}
 C_1(x,x,x, y)=x (x (x  y)) - y \{x,x,x\}, \quad  C_2(x, y, y,x)=2 x \{x, y, y\}+2 \{x, y,x  y\}, \\[2mm]
  \displaystyle  D_1(x,x,x,x, y)=\{x,\{x, y,x\},x\}+\frac{1}{2} \{x, y,\{x,x,x\}\}-\frac{1}{2} x \{x,x  y,x\}.
 \end{array}
\end{equation}
by virtue of the fourth degree identities $Q_1=Q_2=0$. 
\end{lem}
{\Proof} Let us denote the expressions for $C_1, C_2, D_1$ from Lemma \ref{lemma34} by $\tilde C_1, \tilde C_2$ and $\tilde D_1,$ respectively. The statement follows from the following formulas:
$$
C_1(x,x,x, y)-\tilde C_1(x,x,x,y) = \frac{1}{3} \Big(-Q_1(x, x, y, x) + 2 Q_1(x, y, x, x) + 4 Q_1(y, x, x, x) + 
   4 Q_2(x, y, x, x)\Big);
$$
$$
C_2(x,y,y,x)-\tilde C_2(x,y,y,x) = -\frac{1}{3}  Q_2(x, y, y, x);
$$
$$
\begin{array}{c}
\displaystyle D_1(x,x,x,x,y)- \tilde D_1(x,x,x,x,y) = \frac{1}{12} \Big(-10 \,Q_1(x, x, x, x y) - 2 Q_1(x, x,  x y, x) - \\
   5 Q_1(x,  x y, x, x) +  
   3 x\, Q_1(x, y, x, x) - 3  x \,Q_1(y, x, x, x) +  2 \,Q_2(x, x, x,  x y) - 
   2 x \,Q_2(x, y, x, x)\Big).
\end{array}
$$
\subsection{Identities of degree 5, 6 and 7}

Let us define the symmetry \eqref{sym} by formulas \eqref{AA}, \eqref{BBb} and \eqref{CCD}. 
Then one can verify that the compatibility condition \eqref{com} is equaivelent to the following list of identities:
\begin{itemize}
\item  fourth degree identities $Q_1=Q_2=0$; 
\item  fifth degree identities $R_1=R_2=R_3=R_4=R_5=0$;
\item  sexth degree identities $S_1=S_2=S_3=0$;
\item  seventh degree identities $T_1=T_2=0$,
\end{itemize}
where $Q_i$ are defined by \eqref{id4-1}, \eqref{id4-2} and
$$\begin{array}{l}
R_1(x, y) = x (x \{x, y, x\})) + x (y \{x, x, x\})- 
 (x y) \{x, x, x\} - 
 \{x, x (x y), x\}, 
\end{array}
$$
$$\begin{array}{l} 
R_2(x, y) = 4 x ( x (y (x y))) - 4 x ( x \{x, y, y\})+ 
 2 x (y (x ( x y))) - 4 x ( y \{x, x, y\}) + \\[1mm]
	
  8 x \{x, y, x y\} - 2 x \{x, x y, y\} - 
 2 y (x (x (x y))) + 2 y (x \{x, y, x\}) + \\[1mm]
	
 2 y ( y \{x, x, x\}) + 
 4 y \{x, x, x y\} + y \{x, x y, x\} - 
 2 (x y) (x (x y)) + \\[1mm]

 2 \{x, y, \{x, x, y\}\} - 3 \{x, y, \{x, y, x\}\} - 
 2 \{x, y (x y), x\} + \{y, y, \{x, x, x\}\},
\end{array}
$$
$$\begin{array}{l} 
R_3(x, y) =2 x (y \{y, y, y\}) - y \{y, x y, y\} - 
 2 (x y) \{y, y, y\} - \{x, y, \{y, y, y\}\} - \\[1mm]
	
 2 \{y, y, \{x, y, y\}\} + 
 \{y, y, \{y, x, y\}\} + 
 2 \{y, \{x, y, y\}, y\},
\end{array}
$$
$$\begin{array}{l}  
R_4(x, y) =4 x (x (y ( x z))) - 8 x (x \{x, z, y\}) + 
 4 x (y (x (x z))) - 2 x (y \{x, z, x\}) - \\[1mm]
	
 8 x (z \{x, x, y\}) - 2 x (z \{x, y, x\}) + 
 4 x \{x, z, x y\} - 6 x \{x, x z, y\} - \\[1mm]
	
 x \{x, y z, x\} - y \{x, x z, x\} - 
 2 (x y) \{x, z, x\} + 2 (x z) \{x, y, x\} + \\[1mm]
	
 4 (y z) \{x, x, x\} + 
 2 \{x, z, \{x, x, y\}\} - 
 3 \{x, z, \{x, y, x\}\} - 
 4 \{x, x z, x y\} + \\[1mm]
	
 4 \{x, x (y z), x\} + 
 2 \{x, y (x z), x\} - 
 2 \{x, z (x y), x\} + 
 \{y, z, \{x, x, x\}\},
\end{array}
$$
$$\begin{array}{l} 
R_5(x, y) =-4 x (x (z ( y z))) - 4 x (x \{z, y, z\}) - 
 4 x (y \{z, x, z\}) - 4 x (z (x (y z))) + \\[1mm]
	
 4 x (z (z (x y))) + 4 x (z \{x, z, y\}) + 
 6 x \{x, y z, z\} + 4 x \{y, z, x z\} + \\[1mm]
	
 8 x \{z, y, x z\} + 4 x \{z, z, x y\} - 
 2 x \{z, x y, z\} - 2 x \{z, x z, y\} - \\[1mm]
	
 2 y (z \{x, z, x\}) - 4 y \{x, z, x z\} - 
 2 y \{x, x z, z\} - 4 z \{x, x y, z\} - \\[1mm]
	
 2 z \{x, x z, y\} + z \{x, y z, x\} - 
 4 (x y) \{x, z, z\} - 8 (x z) \{x, y, z\} - \\[1mm]
	
 4 (x z) \{x, z, y\} - 2 (y z) \{x, z, x\} - 
 4 \{x, y, \{x, z, z\}\} + 2 \{x, y, \{z, x, z\}\} - \\[1mm]
	
 4 \{x, z, y (x z)\} + 4 \{x, z, z (x y)\} - 
 6 \{x, z, \{x, y, z\}\} + 2 \{x, z, \{x, z, y\}\} + \\[1mm]
	
 2 \{x, z, \{z, x, y\}\} + 4 \{x, y z, x z\} - 
 2 \{x,  z (y z), x\} - 2 \{x, \{z, y, z\}, x\} + \\[1mm]
	
 2 \{y, z, \{x, x, z\}\} - 3 \{y, z, \{x, z, x\}\} + 
 4 \{z, y, \{x, x, z\}\} - 2 \{z, y, \{x, z, x\}\} + \\[1mm]
	
 2 \{z, z, \{x, x, y\}\} - 3 \{z, z, \{x, y, x\}\} + 
 2 \{z, \{x, y, x\}, z\} + 4 \{z, \{x, z, x\}, y\} - \\[1mm]
	
 4 \{x z, z, x y\},
\end{array}
$$
 
$$\begin{array}{l} 

S_1(x,y) = 2 x (x (x \{x, y, x\})) + x (x \{x, x y, x\}) - 
 x \{x, y, \{x, x, x\}\} - x \{x,  x (x y), x\} - \\[1mm]

 2 x \{x, \{x, y, x\}, x\} + 2 \{x, x, x\} \{x, y, x\} + 
 \{x, x y, \{x, x, x\}\} -\\[1mm]
 2 \{x,  x (x(x y)), x\} + 
 2 \{x,  y \{x, x, x\}, x\} + 2 \{x, \{x, x y, x\}, x\},
 
\\[5mm]
 
S_2(x,y) = 2  x (x (x \{y, y, y\})) + x (x \{y, x y, y\}) + 
 2 x (y \{x, x y, y\})+ 2 x \{x, y, \{x, y, y\}\} - \\[1mm]

 x \{x, y, \{y, x, y\}\} - 4 x \{x, \{x, y, y\}, y\} - 
 2 x \{x, \{y, y, y\}, x\} - 2 x \{y, y, \{x, x, y\}\} + \\[1mm]

 x \{y, y, \{x, y, x\}\} - 2 x \{y, \{x, y, x\}, y\} + 
 2 \{x, x, x\} \{y, y, y\} + 2 \{x, y, x\} \{x, y, y\} - \\[1mm]

 2 \{x, y, y \{x, y, x\}\} - 4 \{x, y, \{x, y, x y\}\} - 
 2 \{x, x \{y, y, y\}, x\} - 2 \{x, y \{x, y, y\}, x\} + \\[1mm]

 4 \{x, \{x, y, y\}, x y\} - 2 \{x, \{y, y, x y\}, x\} + 
2 \{x y, y, \{x, y, x\}\},

\end{array}
$$
$$
\begin{array}{l}
S_3(x,y,z) = 4 x ( x (x  \{x, y, z\})) + 8 x ( x ( x \{x, z, y\})) - 
 2 x ( x (z \{x, y, x\})) + 2 x (x \{x, x y, z\})) +\\[1mm]

 4 x (x \{x, x z, y\}) + x (x \{x, y z, x\}) + 
 2 x (y \{x, x z, x\}) + x (z \{x, x y, x\}) - \\[1mm]

 2  x ((x z) \{x, y, x\}) - 2  x \{x, y, \{x, x, z\}\} + 
 3  x \{x, y, \{x, z, x\}\} - 4  x \{x, z, \{x, x, y\}\} -\\[1mm]
 
 2  x \{x, z, \{x, y, x\}\} - 2  x \{x, x y, x z\}\} - 
  x \{x, x (y z), x\} + x \{x, y (x z), x\} - \\[1mm]

 4  x \{x, \{x, y, x\}, z\} - 4  x \{x, \{x, y, z\}, x\} - 
 4  x \{x, \{x, z, x\}, y\} - 8  x \{x, \{x, z, y\}, x\} -\\[1mm]
 
 2  x \{y, z, \{x, x, x\}\} - x \{z, y, \{x, x, x\}\} - 
  z (x \{x, x y, x\}) - 4 z \{x, x, \{x, y, x\}\} + \\[1mm]

  z \{x, y, \{x, x, x\}\} - (x z) \{x, x y, x\} - 
 2  (x (x z)) \{x, y, x\} + 4  \{x, x, x\} \{x, y, z\} + \\[1mm]

 8  \{x, x, x\} \{x, z, y\} - 4 \{x, y, x (x (x z))\}+ 
 4 \{x, y, x \{x, z, x\}\} + 4 \{x, y, z \{x, x, x\}\} + \\[1mm]

 2 \{x, y, \{x, x, x z\}\} + \{x, y, \{x, x z, x\}\} - 
 2 \{x, x (x (y z)), x\} - 2 \{x, x (y (x z)), x\} - \\[1mm]

 4 \{x, x \{x, y, z\}, x\} - 4 \{x, x \{x, z, y\}, x\} + 
 \{x, y z, \{x, x, x\}\} - 2 \{x, y (x (x z)), x\} + \\[1mm]

 4 \{x, z \{x, x, y\}, x\} + 2 \{x, z \{x, y, x\}, x\} + 
 4 \{x, \{x, y, x\}, x z\} + 4 \{x, \{x, z, x\}, x y\} - \\[1mm]

 4 \{x, \{x, z, x y\}, x\} + 2 \{x, \{x, y z, x\}, x\} + 
 \{x z, y, \{x, x, x\}\},
\end{array}
$$
 
$$\begin{array}{l} 
 
 T_1(x,y) = x \{x, x \{x, y, x\}, x\} + 
 \{x, x \{x, x y, x\}, x\} + 
 \{x, \{x, y, x\}, \{x, x, x\}\} - 
\{x, \{x, y, \{x, x, x\}\}, x\},
\end{array}
$$
 
$$\begin{array}{l} 
 
 T_2(x,y) = - 2 x \{x, x y, \{x, y, x\}\} - 2 x \{x, x \{x, y, y\}, x\} + 
 x \{x, y \{x, y, x\}, x]\} - \{x, y, x\} \{x, x y, x\} + \\[1mm]

 2 \{x, y, x \{x, x y, x\}\} + 
 2 \{x, y, \{x, x, \{x, y, x\}\}\} - 
 2 \{x, y, \{x, y, \{x, x, x\}\}\} - \\[1mm]

 3 \{x, y, \{x, \{x, y, x\}, x\}\} + 
 2 \{x,  x \{x, x y, y\}, x\} + 
 \{x,  y \{x, x y, x\}, x\} + \\[1mm]

 4 \{x, \{x, y, x\}, \{x, y, x\}\} + 
 2 \{x, \{x, y, y\}, \{x, x, x\}\} -  
 2 \{x, \{x, y, \{x, x, y\}\}, x\} + \\[1mm]

 3 \{x, \{x, y, \{x, y, x\}\}, x\} - 
 4 \{x, \{x, \{x, y, x\}, y\}, x\} - 
 \{x, \{y, y, \{x, x, x\}\}, x\} + \\[1mm]

 \{\{x, x, x\}, y, \{x, y, x\}\}.
 \end{array}
$$

\begin{thm} All the above identities are equivelent to $Q_1=Q_2={\mathcal R}_4={\mathcal R}_5=0,$
where
\begin{itemize} 
\item ${\mathcal R}_4(x,y,z,u,v)$ is the coefficient of $k_1 k_2 k_3 $ in $R_4(k_1 x + k_2 u + k_3 v, y, z)$; \\
\item ${\mathcal R}_5(x,y,z,u,v)$ is the coefficient of $k_1 k_2 k_3 k_4$ in $R_5(k_1 x + k_2 u, y, k_3 z+k_4 v)$; \\ 
\end{itemize}
and $Q_i$ are defined by \eqref{id4-1},  \eqref{id4-2}.
\end{thm}

\subsection{Jordan triple systems appear}

Identity \eqref{id4-1} can be drastically simplified if we introduce (cf., \eqref{mat2} and \eqref{mat22}) a new triple system $[x,y,z]$ by the formula
\begin{equation}\label{shift}
\{x, y, z\} = [x, y, z]+\frac{1}{2} z (x y) + \frac{1}{2} x (z y).
\end{equation}
Namely, identities \eqref{id4-1} and \eqref{id4-2} become
\begin{equation}\label{iid44-1}
y [u,z,x] - [y u,z,x]- [u,y z,x] - [u,z,y x] = 0,
\end{equation}
and
\begin{equation}\label{iid44-2}
\begin{array}{c}
z \Big(2 [x, u, y] - 2 [x, y, u]- x (u y)\Big) =  {\mathcal J}(z, x,  u y)+ \\[2mm]
 {\mathcal J}(x y, z, u) + 
 {\mathcal J}(y, x z,  u)+ {\mathcal J}(y, z, x u)  - x  {\mathcal J}(y, z, u),
 \end{array}
\end{equation}
where $ {\mathcal J}(x, y, z)$ is defined by \eqref{Jac}.

\begin{thm} The set of identities  $Q_1=Q_2={\mathcal R}_4={\mathcal R}_5=0,$ is equivalent to the identities
  $Q_1=Q_2=0$, and  
\begin{equation}\label{FF}
[x,y,z] =  F(x,z,y) + F(z,x,y)
\end{equation}
for a triple Jordan system $F(x,y,z)$ (cf., \eqref{BB}). 
\end{thm}
Given the triple system $[x,y,z],$ the corresponding triple Jordan system $F(x,y,z)$ can be reconstructed (cf., \eqref{JTS}) by 
\bee\label{JTS1}
F(x,y,z) = \frac{1}{2} \Big( [y,z,x]+[y,x,z]-[x,y,z]  \Big).
\eee

\begin{lem} The identities $Q_1=Q_2=0$ are equivalent to identities \eqref{FId1}, \eqref{FId2}.
\end{lem}

{\bf Resume.}  A system \eqref{mkdv} possesses a symmetry  of the form \eqref{sym} iff the corresponding triple system $B_1$ is defined by a triple Jordan system $F$ by means of \eqref{A1B1}, \eqref{shift}, \eqref{FF} and the skew-symmetric product $A_1$  is related to $B_1$ by \eqref{FId1}, \eqref{FId2},  \eqref{Jac} and \eqref{A1B1}.

\subsection{Lie-Jordan case} 

Consider now the case when the binary operation is a Lie one. Then the identities \eqref{id4-1}, \eqref{id4-2} turn to
\bee
 y \{u,z,x\}- \{y u,z,x\}- \{u,y z,x\}- \{u,z,y x\}=0,\label{id}\\
z \, \Big(\{x,y,u\} - \{x, u,y\} - x (u y) \Big)=0.\label{id1}
\eee

If our system $A$ has no annihllators for the binary operation (that is, if $xA=0$ then $x=0$), then the second identity gives
\bes
\{x,y,u\} - \{x, u,y\} - x (u y)=0,
\ees
 and our system $A$  satisfies identities \eqref{LJ1} - \eqref{LJ5}, that is, $A$ is a Lie-Jordan algebra \cite{GS}. 
Similarly to the case of one operation (binary or trilinear), one can define the notions of reducibility and of irreducible system for the case of two (or more) operations.
An analogue of Lemma \ref{lem_2.1} is true in this case as well. In particular, simple Lie-Jordan algebras correspond to irreducible systems satisfying identities  \eqref{LJ1} - \eqref{LJ5}.  

\begin{lem}
Let $<A,f_1,\ldots,f_k>$ be a vector space $A$ with multilinear operations
$f_1,\ldots,f_k$.
A subspace $I$ of $A$ is called an ideal of the algebraic
system $<A,f_1,\ldots,f_k>$ if for any operation $f_i$ of arity $n_i$ and any
elements $a_1,\ldots,a_{n_i}\in A$ at least one of  which lies in $I$, we have
$f_i(a_1,\ldots,a_{n_i})\in I$.

Assume that $g_1,\ldots,g_m$ are some other multilinear operations on $A$ determined
in terms of operations $f_1,\ldots,f_k$.
Then if $I$ is an ideal of the system  $<A,f_1,\ldots,f_k>$ then it is an ideal of
the system  $<A,g_1,\ldots,g_m>$.
\end{lem}

\Proof
Let $g_j$ be an operation of arity $m_j$ and let $a_1,\ldots,a_{m_j}\in A$  such
that at least one of them lies in $I$. The element $g(a_1,\ldots,a_{m_j})$ is
expressed as a term in operations $f_i$ applied to the elements
$a_1,\ldots,a_{m_j}$. Since $I$ is an ideal with respect to operations $f_i$, the
result should lie in $I$, that is,  $g(a_1,\ldots,a_{m_j})\in I$.

\begin{Cor}\label{Cor3.9}
Let  $<A,f_1,\ldots,f_k>$ and $<A,f_1,\ldots,f_k>$ be two algebraic systems defined
on the same  vector space $A$ such that all operations $f_i$ and $g_j$ are
multilinear. Assume that the operations $g_j$ can be determined as terms in
operations $f_i$ and vice verse, the $f_i$ may be expressed through $g_j$. Then the
system $<A,f_1,\ldots,f_k>$ is simple if and only if so is the system
 $<A,g_1,\ldots,g_m>$.
\end{Cor}
 
 The following theorem describes a structure of simple Lie-Jordan algebras. 
\begin{thm}
Let $L$ be a simple Lie-Jordan algebra. Then $L$ is isomorphic to the Lie-Jordan algebra of skew-symmetric elements $K(A,*)$ for a certain $*$-simple associative algebra with involution $(A,*)$ which is generated by skew-symmetric elements. Conversly, for any $*$-simple associative algebra with involution $(A,*)$ which is generated by skew-symmetric elements, the Lie-Jordan algebra $K(A,*)$ is simple.
\end{thm}
 \Proof
 Let $L$ be a simple Lie-Jordan algebra. By \cite{GS}, there exists an associative algebra with involution $(A,*)$ such that $L$ is isomorphic to the Lie-Jordan algebra $K=K(A,*)$; besides, $A$ is generated by the set $K$. Consider a family of ideals $I$ of the algebra $A$ such that $I^*=I,\ I\cap K=0$. Clearly, this set is inductive, hence by the Zorn Lemma there exists a maximal ideal $I$ in this family. Consider the quotient algebra $\bar A=A/I$. Since $I^*=I$, the algebra $\bar A$ inherits involution $*$. Since $K\cap I=0$, we have $L\cong K(A,*)\cong K(\bar A,*)$. Moreover, for any ideal $I$ of $\bar A$ such that $I^*=I\neq 0$ we have   $I\cap K(\bar A,*)\neq 0$. Clearly, $I\cap K(\bar A,*)$ is an ideal of $K(\bar A,*)$, hence
$I\cap K(\bar A,*)=K(\bar A,*)$ and  $K(\bar A,*)\subseteq I$. Since $\bar A$ is generated by $K(\bar A,*)$, this implies $I=\bar A$. Therefore, the algebra $\bar A$ is $*$-simple. 

Conversly, if $(A,*)$ is a $*$-simple associative algebra with involution $*$ then the triple Jordan system $K(\bar A,*)$ is simple (see, for example, \cite{Zel}), and hence it is simple as a Lie-Jordan algebra.

\ctd

\begin{Cor}
Let $A$ be a simple finite dimensional Lie-Jordan algebra over an algebraically closed field $F$. Then $A$ is isomorphic to one of the following algebras:
\begin{itemize}
\item  $K(M_n,'),\, n>2 $, where $a\mapsto a'$ is the orthogonal involution (the transposition) in $M_n(F)$;
\item  $K(M_{2n},sp), n\geq 1$, where $a\mapsto sp(a)$ is the symplectic involution in $M_{2n}$;
\item  $M_n(F)$, with binary operation $[a,b]=ab-ba$ and ternary operation $\{a,b,c\}=abc+cba$.
\end{itemize}
\end{Cor}
\Proof
It is well known that a $*$-simple algebra $A$ is either a simple algebra or $A\cong B\oplus B^{op}$, with exchanging involution $(a,b)^*=(b,a)$, where $B$ is a simple algebra and  $B^{op}$ is the opposite algebra to $B$. Remind that the opposite algebra $B^{op}$ has the same underlying vector space as $B$, with multiplication $a\cdot b=ba$, where $xy$ stands for multiplication in $B$. 
A simple finite dimensional algebra over an algebraically closed field $F$ is isomorphic to a matrix algebra $M_n(F)$, and it is well known that any involution in it is of orthogonal or symplectic type, which gives us the first two cases. Finally, for the case $A=B\oplus B^{op}$  it is easy to see that the Lie-Jordan algebra $K(A,*)\cong B^{(-)}$, with operations $[a,b]=ab-ba, \ \{a,b,c\}=abc+cba$.

\ctd

\subsection{The case of zero triple Jordan system} In the case of zero binary operation we arrive at the integrable systems described in Section  \ref{sec2}.  Suppose now that $[x,y,z]=0$ for any $x,y,z.$ Then identities \eqref{iid44-1}, \eqref{iid44-2} reduce to
\begin{equation*}\label{iid44-3} 
z \Big(  x (u y)\Big) =  {\mathcal J}(z, x,  u y)+  
 {\mathcal J}(x y, z, u) + 
 {\mathcal J}(y, x z,  u)+ {\mathcal J}(y, z, x u)  - x  {\mathcal J}(y, z, u),
\end{equation*}

Taking here $u=z=x$ we get  $((yx)x)x=0$, therefore our anticommutive operation is 3-engelian.
Since the system 
 is finite dimensional, by \cite{Kuz} the bilinear operation  is nilpotent. 
But a nilpotent algebra can not be simple, hence in this case there are no irreducible systems.

\smallskip
Observe that  there is only one simple Lie-Jordan algebra $A=F$ with trivial binary operation $[x,y]$.

\smallskip

It remains an open question whether there exist non Lie-Jordan simple systems with the Lie binary operation that satisfy identities \eqref{id}, \eqref{id1}.

\section{Acknowledgments} 

The research was initiated during a stay of the second author at the University of S\~ao Paulo, supported by the FAPESP grant 2016/07265-8.
The author  thanks FAPESP for the support and the Institute of Mathematics for providing excellent conditions for the stay. He was also supported by the Russian state assignment No 0033-2019-0006.
The first author was partially supported by FAPESP grant 2014/09310-5 and CNPq grant 303916/2014-1.

\end{document}